\documentclass[twocolumn, pra, showpacs,superscriptaddress]{revtex4}
\usepackage{graphicx}
\usepackage{amsmath}
\usepackage{bm}
\usepackage{float}
\usepackage{color}
\usepackage{sidecap}
\allowdisplaybreaks
\usepackage{soul}
\setstcolor{red}

\begin{document}
\title{Atomic spin-orbit coupling synthesized with magnetic-field-gradient pulses}
\author{Zhi-Fang Xu}
\affiliation{Department of Physics, University of Tokyo, 7-3-1 Hongo, Bunkyo-ku, Tokyo 113-0033, Japan}
\author{Li You}
\affiliation{State Key Laboratory of Low Dimensional Quantum Physics,
Department of Physics, Tsinghua University, Beijing 100084, China}
\author{Masahito Ueda}
\affiliation{Department of Physics, University of Tokyo, 7-3-1 Hongo, Bunkyo-ku, Tokyo 113-0033, Japan}

\date{\today}

\begin{abstract}
We discuss a general scheme for creating atomic spin-orbit coupling (SOC)
such as the Rashba or Dresselhaus types using magnetic-field-gradient
pulses. In contrast to conventional schemes based
on adiabatic center-of-mass motion with atomic internal states
restricted to a dressed-state subspace,
our scheme works for the complete subspace of a hyperfine-spin manifold by utilizing
the coupling between the atomic magnetic moment and external magnetic fields.
A spatially dependent pulsed magnetic field acts as an internal-state-dependent impulse,
thereby coupling the atomic internal spin with its orbital
center-of-mass motion, as in the Einstein-de Haas effect. This effective
coupling can be dynamically manipulated to synthesize SOC of any type (Rashba, Dresselhaus,
or any linear combination thereof).
Our scheme can be realized with most experimental setups of ultracold atoms
and is especially suited for atoms with zero nuclear spins.
\end{abstract}


\maketitle

\section{Introduction}
Synthetic gauge fields recently proposed in the field of quantum gases
are widely perceived as being capable of significantly expanding the scopes
and possibilities of quantum simulations in condensed matter systems \cite{dalibard2011}.
A U(1) abelian gauge field for neutral atoms allows for exploration of
many-body physics such as fractional quantum Hall effects \cite{cooper2008}.
It acts on atomic internal states much like a magnetic field acting on
a charged particle, and has been realized experimentally
in atomic condensates using two Raman lasers \cite{gunter2009,lin2009,fu2011},
in an optical lattice using Raman-assisted tunneling \cite{aidelsburger2011},
and in a driven lattice as well \cite{struck2012}.
The group of Spielman \cite{lin2011}
made an important first step by realizing a spin-orbit coupling (SOC) in
a pseudo spin-1/2 system. Several other groups have also been able to realize
the same form of SOC not only for bosons \cite{zhang2012,qu2013} but also for fermions \cite{wang2012,cheuk2012}.

For a two-dimensional system, one commonly distinguishes between two types of SOC:
the Rashba SOC $p_xF_y-p_yF_x$, which can be transformed into $p_xF_x+p_yF_y$ via a spin rotation,
and the Dresselhaus SOC $p_xF_y+p_yF_x$, where
$p_{x,y}$ are the atomic momenta, while $F_{x,y}$ are the spin-$F$ matrices.
The experiment in Ref. \cite{lin2011} realized a special type of SOC which is
an equally weighted sum of the above two types ($\propto p_xF_y$).
Even richer physics can be simulated with non-abelian gauge fields
of more general forms which create, for example,
the triangular, square, and kagome lattice phases \cite{xu2011,kawakami2011,hu2012,sinha2011,xu2012,ruokokoski2012,xu2013}.
While these phases appear as ground states of spinor Bose-Einstein condensates (BECs),
they need unequal superpositions of Rashba and Dresselhaus SOCs.
Moreover, Majorana fermions can be realized in 2D fermi gases \cite{liu2011,sau2011,liu2012}
in more general forms of SOC.

Several theoretical proposals have discussed how either Rashba or Dresselhaus SOC
can be implemented in neutral atoms \cite{ruseckas2005,sau2011,campbell2011,xu2011b}.
Most of them rely on the idea of adiabatic atomic motion in a subspace spanned by several
spatial-dependent dressed states which are isolated from other levels \cite{wilczek1984,dalibard2011}.
An atom with $N$ internal states is described by
an effective Hamiltonian
\begin{eqnarray}
  H=\sum_i{\mathbf P}_i^2/2m+\sum_{ij}V_{ij},
  \label{hamiltonian}
\end{eqnarray}
where ${\mathbf P}_i$ is the momentum operator associated with the $i$-th internal state
and $V_{ij}$'s denote the internal states' bare energy ($i=j$) and coupling between them ($i\ne j$).
When considering a spatial-dependent unitary
transformation for the internal basis states
$|\Psi_i(\mathbf{r})\rangle=\sum_{j=1}^ND_{ij}(\mathbf{r})|\Phi_j(\bf{r})\rangle$,
the Hamiltonian in the new (possibly adiabatic) basis $|\Phi_i(\mathbf{r})\rangle$
becomes
\begin{eqnarray}
  H'=\sum_{ij}({\bf P}_i\delta_{ij}-{\bf A}_{ij})^2/2m+V'_{ij},
  \label{hamiltonian2}
\end{eqnarray}
with a gauge potential ${\bf A}\equiv i\hbar D^{\dag}({\bf r})\nabla D({\bf r})$
and $V'\equiv D^{\dag}({\bf r})VD({\bf r})$.
The unitary transformation is nontrivial if the matrix $V'$ is approximately block 
diagonalizable and the residual coupling between blocks can be neglected 
due to large energy differences among blocks.
If one of the blocks is spanned
by more than one transformed internal state, non-abelian gauge fields
appear. An effective approach to realize SOC is
to choose a proper transformation $D$.
For instance, in the multipod scheme, laser beams arranged
in a planar form generate a Rashba-type SOC for both
spin-1/2 and spin-1 atoms \cite{ruseckas2005}.
The protocol in Ref.~\cite{campbell2011}
cyclically couples several ground or metastable states with lasers to
overcome collisional decay encountered in the multipod scheme
in which dark states are not the lowest single-particle energy states.
Inspired by the work of Lin {\it et al.} \cite{lin2011},
two protocols capable of synthesizing pure Rashba and Dresselhaus SOC have also been proposed:
one employs a two-dimensional periodic potential formed with
bichromatic laser beams which are retro-reflected along
two orthogonal directions \cite{sau2011},
and the other dynamically generates the two terms of the Rashba SOC
in alternate time intervals \cite{xu2011b}.

Apart from restrictions of their own, none of the above mentioned
proposals can be extended to higher spins.
In this article, we propose a different approach which synthesizes
SOC using space-dependent magnetic field pulses.
As we discuss below, our scheme can readily be implemented
using the currently available cold atom experimental setups
and techniques. It has the potential to overcome
difficulties encountered in previous schemes such as
complicated coupling schemes and
rather specialized experimental systems.
Furthermore, our proposal has the appealing feature of being
extendable to higher spins, and it is especially suitable for atoms with zero nuclear spins
such as $^{52}$Cr \cite{griesmaier2005}, $^{164}$Dy \cite{lu2011},
and $^{168}$Er \cite{aikawa2012}, which do not suffer quadratic Zeeman shifts.

\begin{figure}[tpb]
\centering
\includegraphics[width=\columnwidth]{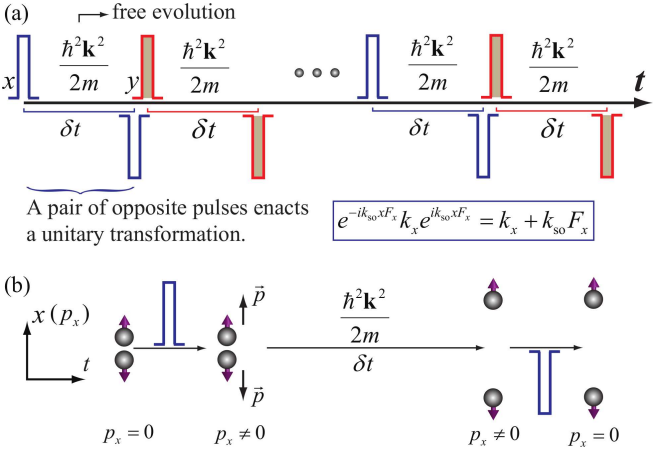}
\caption{(Color online). (a) Schematic illustration of
creating SOC in an arbitary spin-$F$ atom.
A pair of opposite magnetic-field pulses causes a
spin-dependent momentum change.
The pulses along the $x$- ($y$-) direction are denoted in blue
(red-shaded) color.
(b) An illustration for the relative displacement and
momentum of the spatial wave packets for the two spin states.
The spin-dependent impulses lead to a spatial separation of the wave packets
upon free evolution as in a beam splitter.
The thick arrows inside the shaded dots denote the two spin states. }
\label{fig1}
\end{figure}

\section{Protocol for synthesizing SOC}
The SOC we discuss is in general non-abelian; therefore,
it cannot be simply gauged away through a unitary transformation \cite{ho2011}.
The noncommutativity between two operators, e.g. $p_xF_x$ and $p_yF_y$ in a Rashba SOC,
is ubiquitous in quantum systems.
For instance, in a one-dimensional harmonic trap, the two terms in
the Hamiltonian $H=p_x^2/2m+m\omega^2x^2/2$ do not commute
because $[x,p_x]=i\hbar$. Nevertheless, its quantum dynamics can be
numerically simulated using the Trotter expansion $\exp\{-iH\delta t/\hbar\}\simeq
\exp\{-i(p_x^2/2m)\delta t/\hbar\}\exp\{-i(m\omega^2x^2/2)\delta t/\hbar\}$,
to the lowest-order approximation.
Alternatively, the inverse process of effecting two noncommuting terms
in subsequent time intervals can be adopted to realize
an effective dynamics with a Hamiltonian containing noncommuting
operators as in the Rashba or Dresselhaus SOC.
As long as the two noncommuting terms, $k_xF_x$ and $k_yF_y$,
are realized in two different time intervals, simple
unitary transformations can be applied to create them by
making use of the noncommuting nature between position $(x,y)$ and momentum $(p_x,p_y)$.
Each of the unitary transformation requires two space-dependent magnetic field pulses
of opposite signs as illustrated in Figure \ref{fig1}.
For a single cycle, the time evolution operator is given by
\begin{widetext}
\begin{eqnarray}
  U(T,0)&=&\left[U_y(\delta t')
  e^{-i\frac{\hbar\mathbf{k}^2}{2m}\delta t}
  U_y^{\dag}(\delta t')\right]\times
  \left[U_x(\delta t')
  e^{-i\frac{\hbar\mathbf{k}^2}{2m}\delta t}
  U_x^{\dag}(\delta t')\right]
  \nonumber\\
  &=&\exp\left\{-i\frac{\hbar^2}{2m}\left(k_x^2+(k_y+k_{\rm so} F_y)^2 \right)\delta t/\hbar\right\}
  \exp\left\{-i\frac{\hbar^2}{2m}\left(k_y^2+(k_x+k_{\rm so} F_x)^2 \right)\delta t/\hbar\right\}
  \nonumber\\
  &\simeq&\exp\left\{-i\left[\frac{\hbar^2}{2m}(k_x^2+k_y^2)+\frac{\hbar^2 k_{\rm so}}{2m}(k_xF_x+k_yF_y)
    +\frac{\hbar^2k_{\rm so}^2}{4m}(F_x^2+F_y^2)
  \right]2\delta t\right\}\exp\left(\mathcal{O}(\delta t^2)\right),
  \label{timeevolution}
\end{eqnarray}
\end{widetext}
where
\begin{eqnarray}
   U_\epsilon(\delta t')&=&\exp\{-iE_\epsilon'\epsilon F_\epsilon\delta t'/\hbar\},
\end{eqnarray}
and the leading order error (the second order) is estimated to give
\begin{eqnarray}
   \mathcal{O}(\delta t^2)&=&i\frac{\hbar^2k_{\rm so}^2}{4m^2}\delta t^2\{
     2k_xk_yF_z+k_{\rm so}k_y(F_xF_z+F_zF_x)\nonumber\\
     &&+k_{\rm so}k_x(F_yF_z+F_zF_y)+ik_{\rm so}^2[F_y^2,F_x^2]\}.
\end{eqnarray}
We choose $\hbar k_{\rm so}\equiv E'\delta t'$ for $\epsilon=x,y$ by assuming $E'=E_x'=E_y'$
and $T=2\delta t$, where
$E_\epsilon'$ is proportional to the magnetic field gradient $B_\epsilon'$,
which is assumed to be strong enough to satisfy the impulse approximation, whereby
atomic spatial motion during the short pulse interval $\delta t'$ ($\ll \delta t$)
can be neglected. The validity of our protocol requires the neglect of errors
resulting from discrete temporal dynamics from employing the
Trotter expansion. We can reasonably estimate the constraint
on an energy cutoff with the leading-order error term $\mathcal{O}(\delta t^2)$
in Eq.~(\ref{timeevolution}) by enforcing
$\hbar^2k_{\rm so}^2\delta t^2/4m^2\times\max(k_{\rm so}k_{\epsilon},k_xk_y,k_{\epsilon}^2)\ll 1$.
Thus, we realize a Hamiltonian with a Rashba SOC:
$H_R=(p_x^2+p_y^2)/2m+\nu(p_xF_x+p_yF_y)+q(F_x^2+F_y^2)$,
with $\nu=\hbar k_{\rm so}/2m$ and $q=\hbar^2k_{\rm so}^2/4m$ denoting
respectively the strength of SOC and the quadratic Zeeman shift.
More generally, we can realize an arbitrary superposition of the 
Rashba and Dresselhaus SOC using the same
alternating magnetic-field-gradient protocol, but with 
different pulse durations along the x- and y-directions.
For instance, we can take 
$\delta t_{\epsilon}=T|v_{\epsilon}|/\sqrt{v_x^2+v_y^2}$,
the effective time evolution operator becomes
\begin{eqnarray}
  U(T,0)&=&U'_y(\delta t')
  e^{-i\frac{\hbar\mathbf{k}^2}{2m}\delta t_x}
  U_y^{'\dag}(\delta t')\nonumber\\
  &\times&
  U'_x(\delta t')
  e^{-i\frac{\hbar\mathbf{k}^2}{2m}\delta t_y}
  U_x^{'\dag}(\delta t'),
\end{eqnarray}
which gives rise to the an unequal superposition of the
Rashba and Dresselhaus SOC $v_xp_xF_x+v_yp_yF_y$ ($v_{\epsilon}>0$), 
and $U'_{\epsilon}\equiv U_{\epsilon}$. For $v_{\epsilon}<0$,
everything remains the same except that 
$U'_{\epsilon}\equiv U^{\dag}_{\epsilon}$.

A static magnetic field must be divergence-free.
One thus cannot create a magnetic field with only one-directional spatial gradient.
This problem can be circumvented if the other direction with a nonzero gradient
is aligned along the $z$-axis or the direction perpendicular to the 2D planar system
of interest. We further assume that the trapping potential is strongly confined in this direction
to suppress the corresponding atomic center-of-mass motion.
In actual implementation, one can employ a
two-dimensional quadrupole trap (2DQT) \cite{pritchard1983} in the $x$-$z$ plane with $\vec{B}=B'(x,0,-z)$
for the first step of each cycle. For the second step, a second 2DQT in the $y$-$z$ plane
with $\vec{B}=B'(0,y,-z)$ is required. Although the neglect of quadratic Zeeman shifts
causes some error for alkali atoms, such an approximation
becomes exact for $^{52}$Cr \cite{griesmaier2005}, $^{164}$Dy \cite{lu2011}, and $^{168}$Er \cite{aikawa2012}
because their nuclear spins are zero, and hence they have no hyperfine structure.

\begin{figure}[tpb]
\centerline{
\includegraphics[width=3.3in,angle=0]{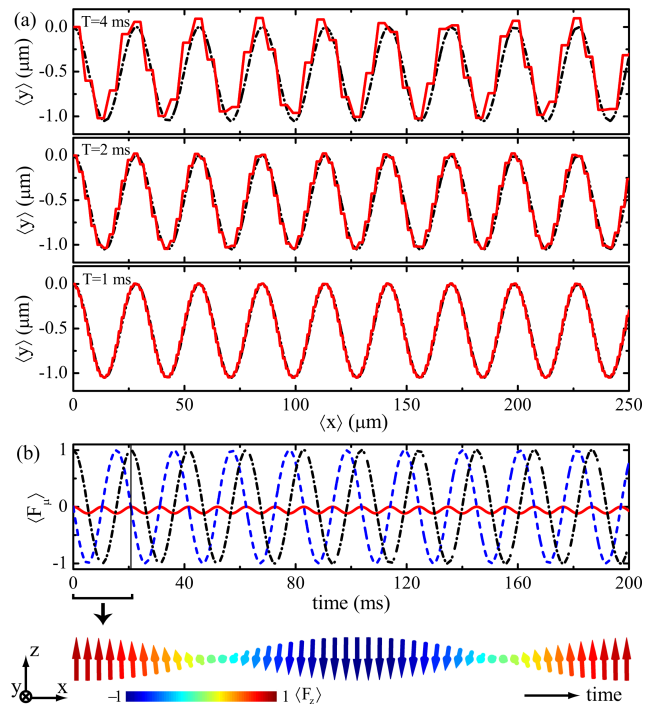}}
\caption{(color online). (a) Semiclassical trajectories for a spin-1 $^{87}$Rb atom in
the $x$-$y$ plane subjected to an effective Rashba-type SOC (black dot-dashed curves)
and to the actual magnetic-field-gradient pulses (red solid curves).
From top to bottom, $T=2\delta t=$ 4, 2, and 1 ms. (b) The corresponding time
evolution of averaged spin components $\langle F_x\rangle$ (red solid curve),
$\langle F_y\rangle$ (blue dashed curve),
and $\langle F_z\rangle$ (black dot-dashed curve) for the semiclassical spatial
motion shown in (a) by black dot-dashed curves. The averaged spin vector referenced
to the space coordinate frame on the lower left corner is shown in the bottom
for a single time period.}
\label{fig2}
\end{figure}

\section{The validity of our protocol}
\subsection{Single-atom motion}
To demonstrate the validity of our protocol in
Eq.~(\ref{timeevolution}), we first compare the
single-atom motion governed by the effective Rashba-type SOC Hamiltonian
$H_R=(p_x^2+p_y^2)/2m+\nu(p_xF_x+p_yF_y)+q(F_x^2+F_y^2)$
with the actual dynamics of repeated magnetic-field-gradient pulses.
The spin-1 $^{87}$Rb atom is used as an example.
At the semiclassical level, the Heisenberg equations of motion for the Hamiltonian $H_R$
are described by
\begin{eqnarray}
  &&\frac{d}{dt}\langle x\rangle=\frac{\langle p_x\rangle}{m}+v\langle F_x\rangle, \quad
  \frac{d}{dt}\langle p_x\rangle=0,
  \nonumber\\
  &&\frac{d}{dt}\langle y\rangle=\frac{\langle p_y\rangle}{m}+v\langle F_y\rangle,
  \quad \frac{d}{dt}\langle p_y\rangle=0,
  \nonumber\\
  &&\frac{d}{dt}\langle F_x\rangle=\frac{v}{\hbar}\langle p_y\rangle\langle F_z\rangle
  +\frac{q}{\hbar}\Big(\langle F_y\rangle\langle F_z\rangle+ \langle F_z\rangle\langle F_y\rangle\Big),
  \nonumber\\
  &&\frac{d}{dt}\langle F_y\rangle=-\frac{v}{\hbar}\langle p_x\rangle\langle F_z\rangle
  -\frac{q}{\hbar}\Big(\langle F_x\rangle\langle F_z\rangle+ \langle F_z\rangle\langle F_x\rangle\Big),
  \nonumber\\
  &&\frac{d}{dt}\langle F_z\rangle=\frac{v}{\hbar}\Big(\langle p_x\rangle\langle F_y\rangle-\langle p_y\rangle\langle F_x\rangle\Big),
  \label{semiclassical}
\end{eqnarray}
where we replace operators by their expectation values,
and the correlations among product operators are ignored.
Initially the atom is at $\langle x\rangle=\langle y\rangle=0$
with its spin fully polarized along the $z$-direction,
i.e., $\langle F_{x,y}\rangle=0$ and $\langle F_z\rangle=1$, $\langle p_y\rangle=0$,
and $\langle p_x\rangle^2/2mk_B=0.01\, (\mu \rm K)$.
For a fixed magnetic field pulse area, or $E'\propto B'\delta t'$,
$\delta t'$ can be shortened by increasing $|B'|$ correspondingly, which
further justifies the impulse approximation. 
According to the Breit-Rabi formula \cite{woodgate}, neglecting the quadratic Zeeman shift 
is quite reasonable in this limit with
$E'=-(5g_I\mu_I+g_J\mu_B)B'/4$, where $g_I$ and $g_J$ are respectively the Land\'e factors for
the nuclear spin ${\mathbf I}$ and the electron with total angular momentum ${\mathbf J}$;
$\mu_I$ and $\mu_B$ are respectively the nuclear magneton and the Bohr magneton.
A conservative estimate gives a rather practical magnetic field gradient 
of $|B'|=50\,\rm G/cm$. For the pulse duration, we take $\delta t'=0.02$ ms
for illustrative purposes. This gives a strength of SOC for
spin-1 $^{87}$Rb atoms characterized by
$k_{\rm so}\simeq 2\pi\times (14\, \rm \mu m)^{-1}$, smaller than 
from the Raman coupling scheme already realized \cite{lin2011}.

Figure \ref{fig2} compares the above effective dynamics with the actual dynamics
for a spin-1 $^{87}$Rb atom in the $x$-$y$ plane with magnetic-field pulses
of duration 4, 2, 1 ms from top to botom. The atomic center-of-mass motion
is neglected during the pulses.
With sufficiently small $\delta t$,
the semiclassical dynamics shows that the atomic trajectories
are essentially identical to those with an effective Rashba-type SOC.
Furthermore, since no trap potential exists in the $x$-$y$ plane,
the trembling motion observed here is analogous to {\it Zitterbewegung},
whose presence in ultracold atoms with SOC was proposed in Refs.~\cite{vaishnav2008,merkl2008},
and the experimental observation was reported in Ref.~\cite{leblanc2013}.

\begin{figure}[tpb]
\centerline{
\includegraphics[width=3.1in]{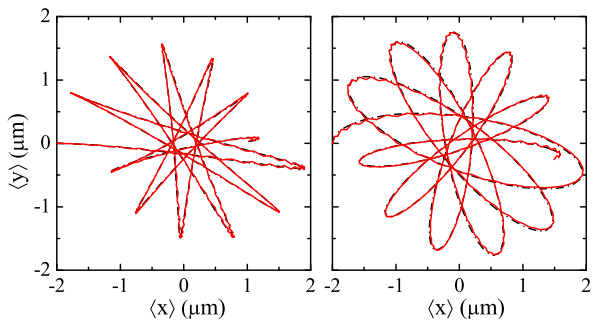}}
\caption{(color online). Quantum trajectories corresponding to Fig. 2 with $T=1$ ms, $x_0=-2\,\mu$m,
$\omega=2\pi\times 30\, \rm Hz$. The spin state is $\zeta=(1,0,0)^T$ for (a) and $\zeta=[-i \cos^2(\pi/8),\sin(\pi/4)/\sqrt{2},i\sin^2(\pi/8)]^T$ for (b).
The trajactories obtained from the effective Rashba-type SOC are shown by black dot-dashed curves. }
\label{fig3}
\end{figure}

Our discussion and derivation above assume a homogenous system.
When an inhomogeneous trapping potential is present,
a similar derivation can be carried out
as long as the unitary transformations from magnetic-field-gradient pulses
commute with local operators. We therefore obtain an effective
Hamiltonian $H'_R=H_R+V(x,y)$, where $V(x,y)$ is the trapping potential.
We now study the quantum motion of a single atom by
numerically solving the corresponding Schr\"odinger equation.
Figure \ref{fig3} shows the numerically calculated trajectories for a
spin-1 $^{87}$Rb atom with an initial off-center Gaussian state
$|\psi\rangle=\zeta\exp\{-((x-x_0)^2+y^2)/2a_{\rm ho}^2\}/\sqrt{\pi}a_{\rm ho}^2$,
where $V(x,y)=m\omega^2(x^2+y^2)/2$,
$a_{\rm ho}=\sqrt{\hbar/m\omega}$, and $\zeta$ is the spin wave function.
We apply the same square magnetic-field-gradient pulses as in
Fig. \ref{fig2}, with the period of each cycle
chosen to be $T=1\,\rm ms$.

Different from the homogenous case,
the atomic linear momentum is no longer conserved when a trapping
potential is present. In the absence of SOC, an atom initially
at rest located at $(x_0,0)$ in the $x$-$y$ plane
will undergo linear oscillations along the $x$-axis.
With SOC, however, atomic center-of-mass motions in the two
orthogonal directions are coupled as a result of the
non-commuting nature between the two vector gauge potentials.
We therefore obtain cyclotron-like motions as shown in Fig. \ref{fig3},
where the actual trajectories depend strongly on the initial spin state.

\begin{figure}[tpb]
\centerline{
\includegraphics[width=3.2in]{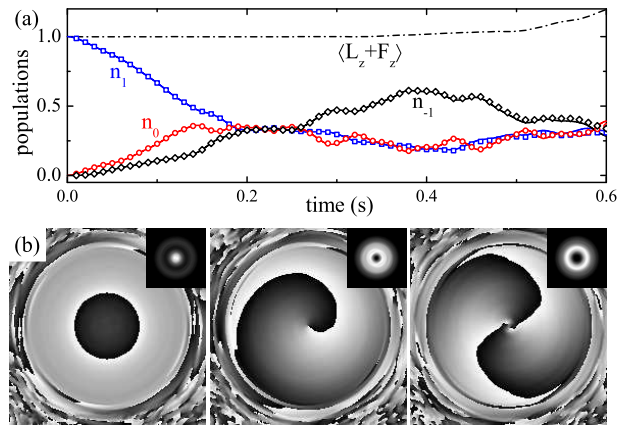}}
\caption{(color online). (a) Time-dependent population fractions $n_{M_F}=N_{M_F}/N$
of a spin-1 $^{87}$Rb condensate with the atom number $N_{M_F}$ in the $M_F$ spin state,
with a Rashba-type SOC (solid lines) or the actual magnetic-field-gradient pulses
for $T=1\,\rm ms$ and $B'\delta t'=1\, \rm G\cdot ms/cm$
(squares, circles, and diamonds for $M_F=1,0,-1$, respectively,
at the end of each 10 periods).
The dot-dashed curve shows the time evolution of the sum of
the $z$-component orbital angular momentum $L_z$
and the spin component $F_z$
from the actual dynamics with magnetic-field-gradient pulses at the end of each period.
(b) Phase distributions of the three spin states ($M_F=1,0,-1$, from left to right, respectively)
at $t = 0.2$ s, obtained using Eq.~(\ref{timeevolution}). Here, the black (white) color
corresponds to the phase $-\pi$ ($\pi$). The corresponding density distributions
over an area of $30 a_{\rm ho}\times 30 a_{\rm ho}$ are shown in the insets.
}
\label{fig4}
\end{figure}

\subsection{Dynamics of a condensate}
Similar to position-dependent
trapping potentials, contact interactions between atoms also
commute with the unitary transformations affected by magnetic-field
pulses. Therefore, the effective Hamiltonian
$H_R$ can be augmented simply by the trapping potential
as well as the interaction terms when a condensate of many atoms is considered.
We can then study the analogous dynamics for a
spin-1 $^{87}$Rb condensate in a pancake potential $V(x,y,z)=m\omega^2(x^2+y^2+\lambda^2 z^2)/2$
with $\omega=2\pi\times 30\, \rm Hz$ and $\lambda=100$, governed by
the effective coupled 2D Gross-Pitaevskii equations
as in Ref.~\cite{xu2010}:
\begin{eqnarray}
    i\hbar\frac{\partial\psi_{\pm1}}{\partial t}&=&
    \left[\frac{}{}H_0+H_{\pm1\pm1}^{\rm ZM}+c_2^{(\rm 2D)}(n_{\pm1}+n_0-n_{\mp1})\right]\psi_{\pm1}\nonumber\\
    &&+c_2^{(\rm 2D)}\psi_{\mp1}^*\psi_0^2+H^{\rm ZM}_{\pm10}\psi_0+H^{\rm ZM}_{\pm1\mp1}\psi_{\mp1},\nonumber\\
    i\hbar\frac{\partial\psi_0}{\partial
    t}&=&\left[\frac{}{}H_0+H_{00}^{\rm ZM}+c_2^{(\rm 2D)}(n_1+n_{-1})\right]\psi_0\nonumber\\
    &&+2c_2^{(\rm 2D)}\psi_0^*\psi_1\psi_{-1}
    +H^{\rm ZM}_{01}\psi_1+H^{\rm ZM}_{0-1}\psi_{-1},
    \label{gpe}
\end{eqnarray}
where $H_0=-\hbar^2\nabla^2/2m+m\omega^2(x^2+y^2)/2$, $n_i=|\psi_i|^2$.
$c_{0,2}^{(\rm 2D)}$ are effective 2D interaction parameters,
and $H^{\rm ZM}$ is the Zeeman term.
We further assume that all atoms are initially populated
in the $M_F=1$ spin state. In Fig. \ref{fig4}(a), we compare the
time-dependent fractional population $n_{M_F}=N_{M_F}/N$ of the $M_F$ spin state,
with the effective Rashba-type SOC (solid curves) to that obtained from the
actual dynamics of the magnetic-field-gradient pulses
with $T=1\,\rm ms$ and $B'\delta t'=1\, \rm G\cdot ms/cm$.
From these comparisons, we conclude that the Hamiltonian with
an effective SOC well describes the dynamics of the condensate affected
by magnetic-field-gradient pulses. The synthesized SOC allows for the realization
of the Einstein-de Haas effect: atoms in the $M_F=1$ state with a
zero angular momentum will be accompanied by vortices with
vorticity $\hbar$ or $2\hbar$ when transferred to the $M_F=0$ or $-1$ state,
while the $z$-component of the total angular momentum is conserved.
This is clearly demonstrated in Fig. \ref{fig4}. From the dot-dashed curve
of Fig. \ref{fig4}(a) we confirm that $\langle L_z+F_z\rangle$
is almost conserved at the end of each cycle.

\section{An alternative protocol}
Finally, we consider magnetic fields with both the $x$- and $y$-dependences.
As pointed out in the previous section, SOC of pure
Rashba or Dresselhaus types cannot simply be eliminated through unitary transformations.
This raises an almost converse question, that is, whether an effective SOC can be created through
unitary transformations to a Hamiltonian without SOC. Surprisingly, the answer is
yes. We illustrate the following operational protocol with
two magnetic field pulses of 2DQT $\vec{B}=B'(x,-y,0)$ in the $x$-$y$ plane,
sandwitched in between the atomic free evolution,
like the two oscillating fields in the Ramsey interferometry.
The time evolution operator for a spin-1 atom now becomes
\begin{widetext}
\begin{eqnarray}
  &&U(t,0)=U_{x,y}(\delta t')\exp\left(-i\frac{\hbar^2\mathbf{k}^2}{2m} t/\hbar \right)
  U_{x,y}^{\dag}(\delta t')
  \nonumber\\
  &&=\exp\left\{-i\frac{\hbar^2 k_{\rm so}^2t}{2m\rho^4\hbar}\left[
  \begin{array}{c}
    \left(k_x{\rho^2}/k_{\rm so}+[x^2+y^2{\rm sinc}(k_{\rm so}\rho)]F_x
    +xy[{\rm sinc} (k_{\rm so}\rho)-1]F_y
    +[y(1-\cos k_{\rm so}\rho)/k_{\rm so}]F_z\right)^2 \\
    +\left(k_y{\rho^2}/k_{\rm so}-xy[{\rm sinc}(k_{\rm so}\rho)-1]F_x
    -[y^2+x^2{\rm sinc}(k_{\rm so}\rho)]F_y
    -[x(1-\cos k_{\rm so}\rho)/k_{\rm so}]F_z \right)^2\\
  \end{array}
  \right]
  \right\},
  \label{timeevolution2}
\end{eqnarray}
\end{widetext}
where $U_{x,y}(\delta t')=\exp[-iE'(xF_x-yF_y)\delta t'/\hbar]$.
We thus find that two magnetic-field-gradient pulses give rise to an effective Hamiltonian containing spatially dependent
non-abelian gauge fields.
In the limit of weak gauge fields and when
$k_{\rm so}\rho\rightarrow 0$ $(\rho=\sqrt{x^2+y^2}\,)$, this Hamiltonian reduces to
\begin{eqnarray}
  H_{\rm eff}=\frac{(p_x-A_x)^2}{2m}+\frac{(p_y-A_y)^2}{2m},
  \label{hamiltonianeff}
\end{eqnarray}
where $A_x=-\hbar(k_{\rm so} F_x+k_{\rm so}^2yF_z/2)$
and $A_y=\hbar(k_{\rm so} F_y+k_{\rm so}^2xF_z/2)$
involve an effective Dresselhaus-type
SOC, giving $\nabla\times\mathbf{A}\ne0$.

In an actual implementation, the magnetic-field pulses do not have
to be perfectly rectangular as we show earlier.
The more important parameter is the area of a pulse.
Therefore, a reasonably smooth temporal profile will be
sufficient. The effective impulse from the pulse
changes the atomic momentum by $\int_0^{\delta t'} E'\, dt$,
where $E'\propto B'$ should be large enough for us to neglect
atomic motion during $\delta t'$, yet small enough to neglect
effects arising from quadratic Zeeman shifts. The above conditions
seem challenging, yet remain within reach of
current experimental setups of cold atoms.
Although the present discussion on the validity of our effective Hamiltonian
containing SOC is at the single-particle or single-mode (condensate) level,
every step along the way for our protocols can be justified more generally
for the many-body spinor atom system with contact interactions and
in a trapping potential.

\section{Conclusions}
In conclusion, we present a scheme of realizing
an effective pure Rashba or Dresselhaus SOC by using
magnetic-field-gradient pulses. Due to the noncommuting property between momentum
and position operators, by applying $x$- and $y$-dependent magnetic-field
pulses alternatively, we can create a pure Rashba-type SOC.
By monitoring the single atom motion at both the semiclassical and quantum levels,
and also the dynamics of a spin-1 condensate,
we confirm that our protocols are valid and practical
within current experimental technology.
The Dresselhaus-type SOC can be realized by applying two pulses from a single
2DQT, where an effective Hamiltonian involving non-abalian gauge fields
approximately emulates the Dresselhaus-type SOC in the limit of small pulse areas.
Our scheme can help develop more general protocols that synthesize gauge
fields of various types. They remain valid for quantum simulation
studies with atomic quantum gases in a variety of settings and apply
to interacting systems as well, as long as the explicit
atomic interaction commutes with unitary transformations
enacted with magnetic-field-gradient pulses.

This work is supported by Grants-in-Aid for Scientific Research (KAKENHI 22340114 and 22103005),
a Global COE Program ``the Physical Sciences Frontier'',
the Photon Frontier Network Program, from MEXT of Japan,
NSFC (No.~91121005 and No.~11004116), the
research program 2010THZO of Tsinghua University, and
MOST 2013CB922000 of the National Key Basic Research Program of China.
Z.F.X. acknowledges support from JSPS (Grant No. 2301327).

{\it Note added:} Our paper was submitted to PRL on 23th Jan, 2013 and transfered to PRA on 29th Apr, 2013.
The idea we present: synthesizing spin-orbit coupling by spatially dependent magnetic fields,
is similar to that of \cite{anderson}. While we restrict atomic motions in a quasi 2D $xy$-plane,
a strong bias along the $z$-axis and a fast oscillating magnetic field along the $x$-$z$ plane is used in \cite{anderson}.
The sinusoidal magnetic field amplitude modulation replaces the two magnetic gradient pulses,
however, the net effect is essentially the same as ours.


\begin{thebibliography}{10}

\bibitem{dalibard2011}
  J. Dalibard, F. Gerbier, G. Juzeliūnas, and P. \"Ohberg,
  Rev. Mod. Phys. \textbf{83}, 1523 (2011).

\bibitem{cooper2008}
  N. R. Cooper, Advances in Physics \textbf{57}, 539 (2008).

\bibitem{gunter2009}
  K. J. G\"unter, M. Cheneau, T. Yefsah, S. P. Rath, and J. Dalibard,
  Phys. Rev. A \textbf{79}, 011604 (2009)

\bibitem{lin2009}
  Y.-J. Lin {\it et al.},
  Phys. Rev. Lett. \textbf{102}, 130401 (2009);
  Y.-J. Lin {\it et al.}, Nature (London) \textbf{462}, 628 (2009).

\bibitem{fu2011}
  Z. Fu {\it et al.}, Phys. Rev. A \textbf{84}, 043609 (2011).

\bibitem{aidelsburger2011}
  M. Aidelsburger {\it et al.}, Phys. Rev. Lett. \textbf{107}, 255301 (2011).

\bibitem{struck2012}
  J. Struck {\it et al.}, Phys. Rev. Lett. \textbf{108}, 225304 (2012).

\bibitem{lin2011}
  Y.-J. Lin, K. Jim\'enez-Garc\'ia, and I. B. Spielman, Nature (London) \textbf{471}, 83 (2011).

\bibitem{zhang2012}
  J.-Y. Zhang {\it et al.}, Phys. Rev. Lett. \textbf{109}, 115301 (2012).

\bibitem{qu2013}
  C. Qu {\it et al.}, e-preprint arXiv:1301.0658.

\bibitem{wang2012}
  P. Wang {\it et al.}, Phys. Rev. Lett. \textbf{109}, 095301 (2012).

\bibitem{cheuk2012}
  L. W. Cheuk {\it et al.}, Phys. Rev. Lett. \textbf{109}, 095302 (2012).

\bibitem{xu2011}
  Z. F. Xu, R. L\"u, and L. You, Phys. Rev. A \textbf{83}, 053602 (2011).

\bibitem{kawakami2011}
  T. Kawakami, T. Mizushima, and K. Machida,
  Phys. Rev. A \textbf{84}, 011607(R) (2011).

\bibitem{hu2012}
  H. Hu, B. Ramachandhran, H. Pu, and X.-J. Liu,
  Phys. Rev. Lett. \textbf{108}, 010402 (2012).

\bibitem{sinha2011}
  S. Sinha, R. Nath, and L. Santos, Phys. Rev. Lett. \textbf{107}, 270401
  (2011).

\bibitem{xu2012}
  Z. F. Xu, Y. Kawaguchi, L. You, and M. Ueda, Phys. Rev. A \textbf{86}, 033628 (2012).

\bibitem{ruokokoski2012}
  E. Ruokokoski, J. A. M. Huhtam\"aki, and M. M\"ott\"onen, Phys. Rev. A \textbf{86}, 051607 (2012).

\bibitem{xu2013}
  Z. F. Xu, S. Kobayashi, and M. Ueda, e-preprint arXiv:1304.4340.

\bibitem{liu2011}
  J. Liu {\it et al.},
  Phys. Rev. Lett. \textbf{107}, 026405 (2011).

\bibitem{sau2011}
  J. D. Sau {\it et al.},
  Phys. Rev. B \textbf{83}, 140510(R) (2011).

\bibitem{liu2012}
  X.-J. Liu, L. Jiang, H. Pu, and H. Hu,
  Phys. Rev. A \textbf{85}, 021603(R) (2012).

\bibitem{ruseckas2005}
  J. Ruseckas {\it et al.},
  Phys. Rev. Lett. \textbf{95}, 010404 (2005);
  G. Juzeli\=unas, J. Ruseckas, and J. Dalibard,
  Phys. Rev. A \textbf{81}, 053403 (2010).

\bibitem{campbell2011}
  D. L. Campbell, G. Juzeli\=unas, and I. B. Spielman, Phys. Rev. A \textbf{84}, 025602 (2011).

\bibitem{xu2011b}
  Z. F. Xu and L. You, Phys. Rev. A \textbf{85}, 043605 (2012).

\bibitem{wilczek1984}
  F. Wilczek and A. Zee, Phys. Rev. Lett. \textbf{52}, 2111 (1984).

\bibitem{griesmaier2005}
  A. Griesmaier {\it et al.}, Phys. Rev. Lett. \textbf{94}, 160401 (2005).

\bibitem{lu2011}
  M. Lu {\it et al.},  Phys. Rev. Lett. \textbf{107}, 190401 (2011).

\bibitem{aikawa2012}
  K. Aikawa {\it et al.}, Phys. Rev. Lett. \textbf{108}, 210401 (2012).

\bibitem{ho2011}
  T.-L. Ho and S. Zhang, Phys. Rev. Lett. \textbf{107}, 150403 (2011).

\bibitem{pritchard1983}
  D. E. Pritchard, Phys. Rev. Lett. \textbf{51}, 1336 (1983).

\bibitem{woodgate}
  G. K. Woodgate, {\it Elementary Atomic Structure}, Oxford University Press Inc., second edition, 1980.

\bibitem{vaishnav2008}
  J. Y. Vaishnav and C. W. Clark, Phys. Rev. Lett. \textbf{100},153002 (2008).

\bibitem{merkl2008}
  M. Merkl {\it et al.}, Europhys. Lett. \textbf{83}, 54002 (2008).

\bibitem{leblanc2013}
  L. J. LeBlanc, M. C. Beeler, K. Jimenez-Garcia, A. R. Perry, S. Sugawa, R. A. Williams, and I. B. Spielman,
  e-preprint arXiv:1303.0914.

\bibitem{xu2010}
  Z. F. Xu, P. Zhang, R. L\"u, and L. You, Phys. Rev. A \textbf{81}, 053619 (2010).

\bibitem{anderson}
  B. M. Anderson, I. B. Spielman, G. Juzeli\=unas, e-preprint arXiv:1306.2606.

\end{thebibliography}
\end{document}